\documentclass{article}

\usepackage{microtype}
\usepackage{graphicx}
\usepackage{subfigure}
\usepackage{booktabs} 

\usepackage{hyperref}

\usepackage[accepted]{icml2019}

\icmltitlerunning{Unsupervised Temporal Clustering to Monitor the Performance of Alternative Fueling Infrastructure}

\begin{document}

\twocolumn[
\icmltitle{Unsupervised Temporal Clustering to Monitor the \\
           Performance of Alternative Fueling Infrastructure}




\begin{icmlauthorlist}
\icmlauthor{Kalai Ramea}{to}
\end{icmlauthorlist}

\icmlaffiliation{to}{Palo Alto Research Center, Palo Alto, California, USA}

\icmlcorrespondingauthor{Kalai Ramea}{kramea@parc.com}

\icmlkeywords{Climate Change, Transportation, Zero-Emission Vehicles, Clustering, Time-Series, Emissions, Machine Learning}

\vskip 0.3in
]



\printAffiliationsAndNotice{}  

\begin{abstract}
Zero Emission Vehicles (ZEV) play an important role in the decarbonization of the transportation sector. For a wider adoption of ZEVs, providing a reliable infrastructure is critical. We present a machine learning approach that uses unsupervised temporal clustering algorithm along with survey analysis to determine infrastructure performance and reliability of alternative fuels. We illustrate this approach for the hydrogen fueling stations in California, but this can be generalized for other regions and fuels. 

\end{abstract}

\section{Introduction}
\label{intro}

The transportation sector accounts for about 33\% of the total end-use carbon emissions highest among all the sectors in the US. Specifically, emissions from light-duty vehicles account for more than half of the total emissions in the transportation sector \cite{aeoenergyuse}. Several countries have been adopting an increasing number of incentives to decarbonize the light-duty vehicle sector through zero-emission vehicles (ZEV). This has been reflected in the recent uptake of battery electric vehicle (BEV) fleet all over the world \cite{icctWhitepaper}. However, there are still concerns among consumers about purchasing a BEV regarding range limitation, long charging times and charger congestion \cite{chargercongestion}.

To further increase the total market penetration of ZEVs in the light-duty fleet, a portfolio approach of technological diversity is needed that addresses the variety of concerns expressed by the consumers. In California and Japan, hydrogen fuel cell vehicles (FCV) were introduced in recent years and are being increasingly adopted alongside BEVs. There are currently over 5,800 FCVs in California \cite{cafcpnum}, and it is estimated that about 40,000 FCVs will be on the Californian roads by 2022 \cite{arbfuelcell2018}. FCVs have a longer driving range, significantly shorter refueling time (in the order of minutes) and similar fueling mechanism as gasoline. Thus, they can play an important role in expanding the light-duty vehicle ZEV market as they have the potential to complement and alleviate some of the concerns raised about the BEVs.

One of the biggest barriers to FCV adoption is the reliability and availability of hydrogen fueling stations. Unlike BEVs, which have multiple modes of charging (home, work, public chargers), FCVs solely rely on access to public fueling infrastructure. The reputation of an unreliable station may discourage consumers from adopting the technology \cite{Kelley2018}. Therefore, for significant market acceptance of FCVs, policymakers should not only consider a good network of hydrogen refueling stations but also have a metric to assess their performance on a real-time basis.

\textbf{Existing Literature:} In the past, researchers have developed various designs of what constitutes an optimal network of hydrogen stations in the pre-planning stage \cite{Ogden:2011da,StephensRomero2010SystematicPT}. Several modeling approaches have been explored to understand the spread and quantity of hydrogen refueling station designs \cite{Lim2010HeuristicAF,NicholasOgden2007,Honma2008AMM,Bhatti2015AlternativeFS,Kang2015StrategicHR}. However, there has been little focus on assessing and monitoring the performance of existing stations. Although National Renewable Energy Laboratory has published specific data products on hydrogen station infrastructure \cite{nrel}, most of this data explains the technical details of the hydrogen fuel delivery and the adoption of fuel at the regional level on the supply side.

In this paper, we demonstrate an unsupervised temporal clustering approach on the hourly utilization data that was collected as a part of this project. To better understand the reasons behind the clustered categories, we conducted a survey of 100 FCV drivers in California. This approach identifies the stations that are performing within the healthy range, and those that are over-stressed and intervention is needed. We are publicly releasing the hourly station capacity dataset we collected for this research project\footnote{Github link to the dataset will be released if the paper is accepted.}. To the best of our knowledge, this is the first study that focuses on deploying a machine learning algorithm to analyze the micro-level usage of hydrogen fueling stations as a synchronous network to monitor their performance.

\section{Methodology}
\label{method}

The project was completed in three phases as follows: (a) Data collection, (b) Unsupervised temporal clustering, and (c) Survey analysis.

\subsection{Data Collection and Pre-processing}
\begin{figure}[ht]
\vskip 0.2in
\begin{center}
\centerline{\includegraphics[width=\columnwidth]{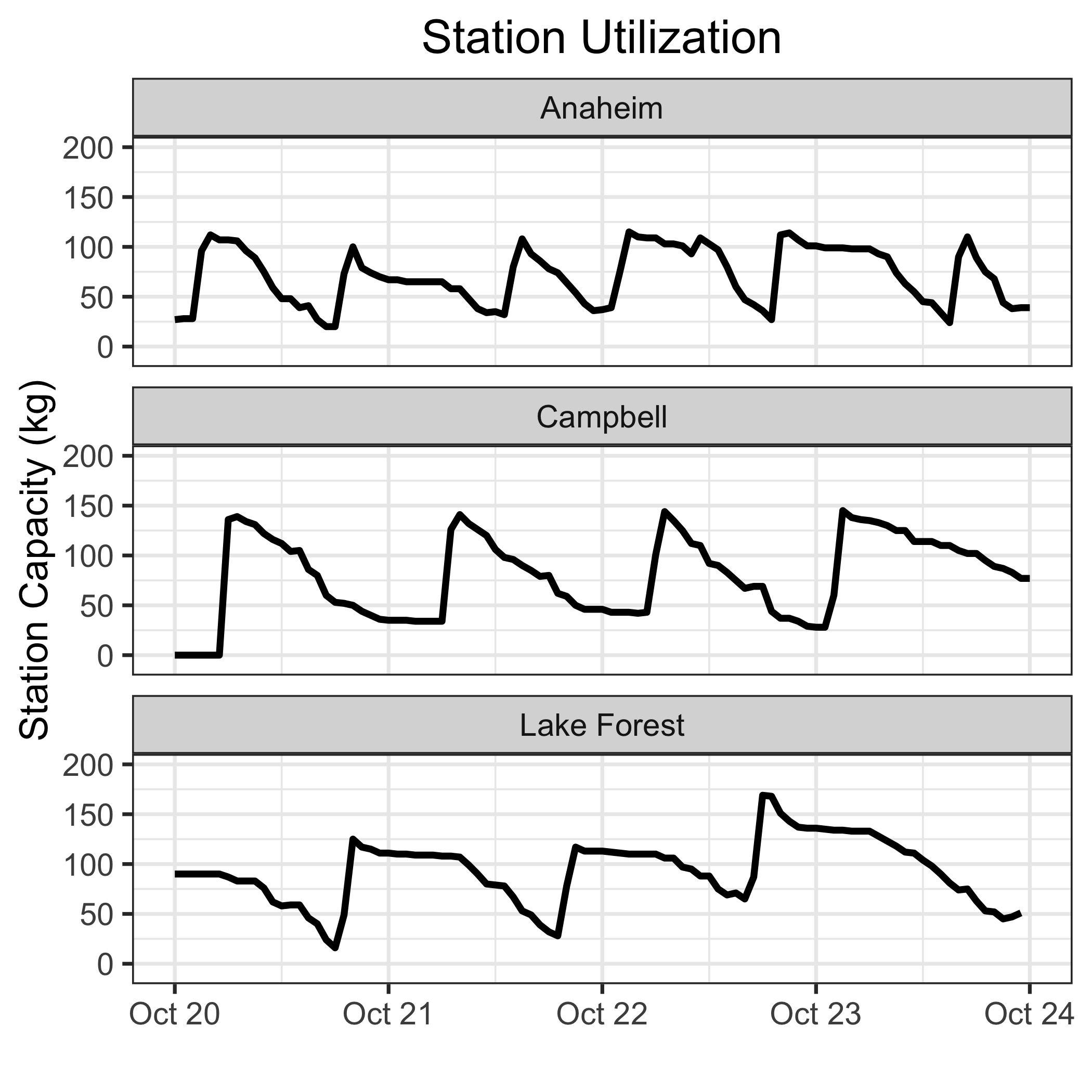}}
\caption{Snapshot of Hydrogen Fuel Utilization for Anaheim, Campbell and Lake Forest stations}
\label{util_time_series}
\end{center}
\vskip -0.2in
\end{figure}

To ground the analysis in real world data, we collected the hydrogen capacity data of all the fueling stations in California. The California Fuel Cell Partnership website \cite{cafcpstn} provides information on the hydrogen fueling station locations and their real-time capacity levels in kg. The results shown in this paper is based on the hourly data collected for a period of three months starting from October 2018 to December 2018. A snapshot of utilization patterns of the stations is shown in Figure \ref{util_time_series}.

The missing data in the time series due to technical issues in reporting were imputed through linear interpolation \footnote{Newport Beach and Torrance stations do not report capacity values on the website, so these stations are ignored in the analysis. Three new stations were added towards the end of December 2018. They were not included in the data collection or analysis.}. As the capacity varies across the stations, the time-series data is normalized before cluster analysis is performed. 

\subsection{Unsupervised Temporal Clustering}

Unsupervised cluster analysis has been extensively used in several domains to study categories of distinct classes occurring in the dataset, with the most common and popular algorithm being the K-Means Clustering \cite{Stuartkmeans}. While this may work well on most univariate or multivariate datasets, this density-based clustering approach has limitations when applied to time-series or spatial data as it cannot handle the dependencies across data points. 

Temporal clustering is increasingly being used in finance, IoT and energy domains to identify behavior variation and anomaly detection \cite{Urosevic, doi:10.1080/14697688.2016.1171378}. Calculating distance measures across multiple time-series sequences is a tricky task, and can often depend on what invariance is considered to measure in the respective domains. However, Batista et al. proved that the complexity invariant distance measure is generally applicable to most time-series datasets \cite{Batista2011ACD}. This distance measure calculates Euclidean distance between the two time-series and a correction factor is applied that factors in the differences in complexities. We use this approach to analyze the behavioral variation across the time-series fuel utilization data of hydrogen stations. 

Although there is an underlying spatial component in the data, the station locations are sparse enough at this point and are considered independent of each other. We confirmed this by calculating \textit{Moran's I coefficient} \cite{Anselin2006GeoDaAI} for the locations at different time intervals, and found that there is no significant spatial autocorrelation. However, as more stations are added in the future, this approach would require a spatiotemporal clustering algorithm to better understand how the location of the station contributes to their respective behavior and utilization.

\subsection{Survey Analysis}

In the third phase of the project, we designed a short, simple survey on hydrogen fueling station performance. We recorded answers from about 100 participants who currently drive FCVs in California\footnote{Survey was hosted on https://www.surveymonkey.com/ website and the results were collected between the 5\textsuperscript{th} and 15\textsuperscript{th} of January, 2019.}. 

The following are the list of questions answered by the survey participants: (a) location, (b) most and least preferred stations, (c) reasons for station preferences, (d) backup stations, and (e) reasons for station avoidance. For the reasons behind the station preferences, respondents were asked to choose all the reasons that apply, such as proximity to home, proximity to work, reliability, hours of operation, safe neighborhood, no station congestion, ease of use, and were asked to check whichever ones apply to them. They were also encouraged to record any other comments under \lq{Other (specify)}\rq).

\section{Results}
\label{results}

\begin{figure}[H]
\vskip 0.2in
\begin{center}
\centerline{\includegraphics[width=\columnwidth]{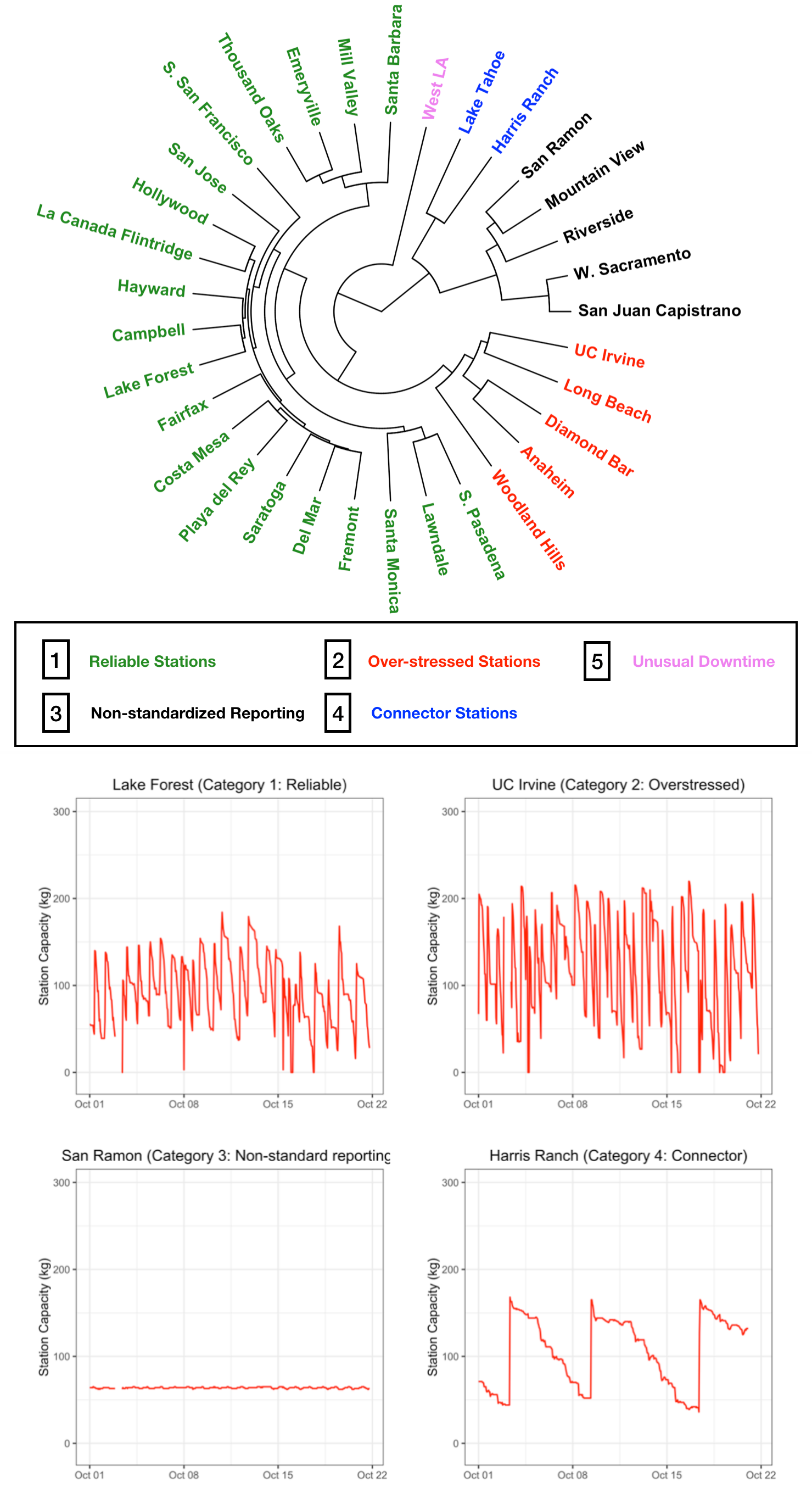}}
\caption{Visualization of the Temporal Cluster Analysis of Hydrogen Fueling Station Utilization}
\label{clusterviz}
\end{center}
\vskip -0.2in
\end{figure}

The results shown in Figure \ref{clusterviz} indicate that there are four significant clusters based on the temporal utilization patterns of hydrogen stations. To understand what these clusters mean, we took the survey responses as well as station attributes to divide them into the following categories.

\begin{enumerate}
    \item{\textbf{Reliable stations:} The top preferred stations identified by the survey respondents as most reliable appear in cluster group \textbf{1}. These stations act as a guideline for determining whether the fueling station is performing as intended.}
    \item{\textbf{Over-stressed stations:} The survey respondents were asked to identify the stations they found to be \lq{too busy}\rq. As cluster group \textbf{2} predominantly aligns with this answer, they are identified as stations that need intervention in terms of increase in storage capacity or a supplementary station nearby.}
    \item {\textbf{Non-standardized Reporting:} Cluster group \textbf{3} of hydrogen stations appear to have a different storage mechanism. The liquid hydrogen fuel is pumped and stored in a bigger tank which is held at a cryogenic temperature, which is then vaporized, compressed and stored in a smaller tank for dispensing \cite{cah2deliverstorage}. The capacity values reported are that of the smaller tank which does not reflect the actual capacity of the station. For standard comparison across stations, this reporting mechanism needs be adjusted.  }
    \item{\textbf{Connector stations:} Two stations, Harris Ranch and Lake Tahoe are clustered together in group \textbf{4} as these are buffer stations for people traveling long distances and are not used on a regular basis.}
    
     \item{\textbf{Unusual downtime:} West LA station stands out in the cluster group as there was an unusual downtime for a number of days during the time of data collection. This is not always the case during other time periods. }
\end{enumerate}

\section{Conclusion}
\label{conclusion}

Adoption of a zero-emission vehicle fleet is a critical component to reducing carbon emissions in the transportation sector. We need diversity of ZEV technologies in the fleet to satisfy the concerns of different segments of consumers. In the case of FCVs, consumers need to trust that they would have reliable fueling infrastructure before they purchase the vehicle. Monitoring of station performance would, therefore, provide transparency to customers and help policymakers to fix issues as they arise.

We show the use of unsupervised temporal clustering algorithm to determine hydrogen refueling station performance based on station capacity time-series data. The survey responses we obtained helped to understand the reasons behind our cluster categories. The analysis presented in this paper is for a static snapshot of time, hence, a machine learning algorithm is not strictly necessary as we could have obtained the same results through a heuristic. However, California plans on expanding the number of stations to about 200 stations in the year 2025, which increases the spatial and temporal complexity of the data. Therefore, a simple heuristic may fail to capture the utilization of these stations and a machine learning algorithm demonstrated in this paper may be necessary.

Even though the analysis presented in this paper focuses on a static snapshot of time, this approach could be implemented as a continuous monitoring effort with real-time streaming of station capacity data. Finally, we are releasing the hourly station capacity dataset that we collected for the machine learning researchers to further explore temporal and spatiotemporal projects. 

\nocite{langley00}

\bibliography{example_paper}
\bibliographystyle{icml2019}

\end{document}